\documentclass[twocolumn]{article}
\begin{document}

\title{
Efficient tests for experimental quantum gates
}
\author{Holger F. Hofmann
\\
JST-CREST, Graduate School of Advanced Sciences of Matter, Hiroshima University\\
Kagamiyama 1-3-1, Higashi Hiroshima 739-8530, Japan\\
TEL: +81-82-424-7652 \quad FAX: +81-82-424-7649\\
e-mail: h.hofmann@osa.org
}

\date{}

\maketitle

\abstract{
Realistic quantum gates operate at non-vanishing noise levels.
Therefore, it is necessary to evaluate the performance of each
device according to some experimentally observable criteria
of device performance. In this presentation, the characteristic
properties of quantum operations are discussed and efficient
measurement strategies are proposed.
}

\section{Introduction}

The technological foundation of quantum information is the control
of physical systems at the quantum level. Within the standard theoretical
framework, this level of control is represented by pure states and by
unitary transformations. However, it is a rather peculiar feature of
quantum theory that these elements of the theory cannot be identified
directly with the effects actually observed in the laboratory.
Instead, experimental observations are connected to the theoretical
formalism using the statistical interpretation given by the measurement
postulate. This indirect connection between experimental evidence and
theoretical interpretation makes it impossible to define the success
or failure of a quantum operation in conventional terms. Specifically,
the result of an individual measurement cannot be used to identify
the quantum state, so it is never possible to tell what the output state
of a quantum process "really" was and how the process actually changed
the input state. The evaluation of noisy quantum operations is therefore
a highly non-trivial task requiring a more detailed analysis of the
connection between the quantum formalism and the observable measurement
statistics \cite{Chu97,Poy97,Bri04,Gil04}.

From the theoretical side, this problem has been addressed by defining
mathematical measures of the distances between states and operations
in their respective vector spaces. In particular, the effective
overlap between two processes can then be given in terms of the
process fidelity $F$, obtained from the trace of the process matrix
products \cite{Bri04,Gil04}. However, it is not immediately clear
how the process fidelity relates to individual tests of quantum gate performance
in the laboratory.

In this presentation, the problem is therefore approached from the
opposite viewpoint: first, the quantum operation is defined in terms
of observable effects, then the classical fidelities of these
properties are related to the overall process fidelity.
In particular, it is pointed out that only two classical operations
can already provide a good measure of how well a quantum gate works
\cite{Hof04}. Possible selections of classical fidelities are considered
for the case of the quantum controlled-NOT, and the problem of verifying
entanglement generation by local measurements is addressed.

\section{Characteristic observable operations
and their classical fidelities}

In general, the effect of an ideal deterministic operation on any
quantum state $\mid \psi \rangle$ is described by the unitary
operator $\hat{U}_0$. However, this unitary operation is actually
a very compact summary of many possible operations that can be
performed by a quantum device depending on the choice of input
states. In order to verify the actual properties of an experimental
quantum process, it is therefore necessary to select a representative
set of observable operations by choosing appropriate input states
and output measurements.

For a set of $N$ distinguishable (= orthogonal) quantum states
$\{\mid n \rangle \}$ of an $N$-level system, the effect of the
unitary operation $\hat{U}_0$ is given by
\begin{equation}
\label{eq:basic}
\hat{U}_0 \mid n \rangle = \mid f_n \rangle.
\end{equation}
This transition from the set of distinguishable states
$\{\mid n \rangle \}$ in the input to the corresponding
distinguishable output states $\{\mid f_n \rangle \}$ is
a characteristic observable operation of the quantum device.
It can be verified by performing an appropriate von Neumann
measurement of the output, with a fidelity of $F_{n \to f_n}$
obtained by averaging the probability of obtaining the correct
output result $f_n$ over all inputs $n$,
\begin{equation}
F_{n \to f_n} = \frac{1}{N} \sum_n p(f_n | n).
\end{equation}
Since the operation transforming $n$ into $f_n$ can be defined
in classical terms, that is, without any reference to the
unobserved effects of quantum coherence, it will be referred to
as a classical fidelity in the following, to distinguish it
from the quantum mechanical concepts of fidelity defined as
measures in Hilbert space.

In the case of a quantum controlled-NOT, one of the characteristic
observable operations is the classical controlled-NOT operation
observed in the computational basis,
\begin{eqnarray}
\hat{U}_{\mbox{CNOT}} \mid 0_z;0_z \rangle &=& \mid 0_z;0_z \rangle
\nonumber \\
\hat{U}_{\mbox{CNOT}} \mid 0_z;1_z \rangle &=& \mid 0_z;1_z \rangle
\nonumber \\
\hat{U}_{\mbox{CNOT}} \mid 1_z;0_z \rangle &=& \mid 1_z;1_z \rangle
\nonumber \\
\hat{U}_{\mbox{CNOT}} \mid 1_z;1_z \rangle &=& \mid 1_z;0_z \rangle.
\end{eqnarray}
The classical fidelity of this operation, $F_{zz \to zz}$, is
obtained by averaging the probability of obtaining the correct
measurement outcome in the $Z$-basis over the four $Z$-basis
input states,
\begin{eqnarray}
F_{zz \to zz} &=& \frac{1}{4}\big( p_{zz|zz}(00|00) + p_{zz|zz}(01|01)
\nonumber \\
&& \hspace{0.5cm} + \; p_{zz|zz}(11|10) + p_{zz|zz}(10|11) \big).
\nonumber \\
\end{eqnarray}
However, the quantum coherence of the gate implies that a characteristic
observable operation exists for any choice of input basis.
In the case of a two qubit gate such as the quantum controlled-NOT,
it is therefore interesting to consider input states taken from the
$X$ and $Y$ bases as well. It is then possible to define a set of
nine characteristic observable operations of the two qubit gate.

An overview over the characteristic operations for the quantum controlled-NOT
is shown in table \ref{qcnot}. Like all unitary operations, the quantum
controlled-NOT has a set of eigenstates for which the characteristic operation
is the identity operation ("identity" in the table).
As can be seen from the table, this set of
eigenstates is the $ZX$ basis. The regular controlled-NOT operation ("CNOT"
in the table) is observable in both the $ZZ$ and the $ZY$ basis, since
the conditional
spin flip of the target qubit corresponds to a rotation around the $X$ axis.
If the target qubit input is an eigenstate of $X$, its eigenvalue is
preserved and the effect on the control bit is a conditional rotation
around the $Z$ axis. Thus, the gate performs a controlled-NOT operations
with reversed roles of the target and the control ("reverse CNOT"
in the table), e.g. for the $XX$ basis,
\begin{eqnarray}
\hat{U}_{\mbox{CNOT}} \mid 0_x;0_x \rangle &=& \mid 0_x;0_x \rangle
\nonumber \\
\hat{U}_{\mbox{CNOT}} \mid 0_x;1_x \rangle &=& \mid 1_x;1_x \rangle
\nonumber \\
\hat{U}_{\mbox{CNOT}} \mid 1_x;0_x \rangle &=& \mid 1_x;0_x \rangle
\nonumber \\
\hat{U}_{\mbox{CNOT}} \mid 1_x;1_x \rangle &=& \mid 0_x;1_x \rangle.
\end{eqnarray}
The same observable operation is obtained for the $YX$ basis.
Finally, the entanglement generating functions of the gate are
described by the observable operations on input states of the
$XY$, $XZ$, $YY$ and $YZ$ bases ("entangle" in the table).
In these four cases, the output
states form a complete Bell basis of four orthogonal maximally
entangled states each, e.g. for the $XZ$ basis,
\begin{eqnarray}
\label{eq:entangle}
\hat{U}_{\mbox{CNOT}} \mid 0_x;0_z \rangle &=& \frac{1}{\sqrt{2}}
\left(\mid 0_z;0_z \rangle + \mid 1_z;1_z \rangle \right)
\nonumber \\
\hat{U}_{\mbox{CNOT}} \mid 0_x;1_z \rangle &=& \frac{1}{\sqrt{2}}
\left(\mid 0_z;1_z \rangle + \mid 1_z;0_z \rangle \right)
\nonumber \\
\hat{U}_{\mbox{CNOT}} \mid 1_x;0_z \rangle &=& \frac{1}{\sqrt{2}}
\left(\mid 0_z;0_z \rangle - \mid 1_z;1_z \rangle \right)
\nonumber \\
\hat{U}_{\mbox{CNOT}} \mid 1_x;1_z \rangle &=& \frac{1}{\sqrt{2}}
\left(\mid 0_z;1_z \rangle - \mid 1_z;0_z \rangle \right).
\nonumber \\
\end{eqnarray}
It should be noted that the direct experimental verification of
this operation requires a non-local Bell measurement of the two
qubits. Thus, characteristic observable operations are not necessarily
local, and it may be useful to distinguish classical local fidelities
such as the identity or the controlled-NOT operations from the
classical non-local fidelities of entanglement generation. Obviously,
the use of the term "classical" in this context is merely based
on the absence of coherent superpositions due to the restriction
of the operation to well-defined basis sets in the input and the
output, not on any problems of separability.

\begin{table}
\begin{picture}(226,230)
\thicklines

\put(0,10){\line(0,1){190}}
\put(55,10){\line(0,1){210}}
\put(112,10){\line(0,1){170}}
\put(167,10){\line(0,1){170}}
\put(226,10){\line(0,1){210}}

\put(55,220){\line(1,0){171}}
\put(0,200){\line(1,0){55}}
\put(30,160){\line(1,0){196}}
\put(30,110){\line(1,0){196}}
\put(30,60){\line(1,0){196}}
\put(0,10){\line(1,0){226}}

\put(0,180){\makebox(55,15){\Large Target}}
\put(0,165){\makebox(55,15){\Large Input}}

\put(0,125){\makebox(55,20){\Large $X$-basis}}
\put(0,75){\makebox(55,20){\Large $Y$-basis}}
\put(0,25){\makebox(55,20){\Large $Z$-basis}}

\put(55,195){\makebox(171,15){\Large Control Input}}

\put(55,170){\makebox(57,20){\Large $X$-basis}}
\put(112,170){\makebox(57,20){\Large $Y$-basis}}
\put(169,170){\makebox(57,20){\Large $Z$-basis}}

\put(55,135){\makebox(57,15){\Large reverse}}
\put(55,120){\makebox(57,15){\Large CNOT}}
\put(112,135){\makebox(57,15){\Large reverse}}
\put(112,120){\makebox(57,15){\Large CNOT}}
\put(169,125){\makebox(57,20){\Large identity}}

\put(55,75){\makebox(57,15){\Large entangle}}
\put(112,75){\makebox(57,15){\Large entangle}}
\put(169,75){\makebox(57,20){\Large CNOT}}

\put(55,25){\makebox(57,15){\Large entangle}}
\put(112,25){\makebox(57,15){\Large entangle}}
\put(169,25){\makebox(57,20){\Large CNOT}}

\end{picture}
\caption{\label{qcnot}
Characteristic operations of the quantum controlled-NOT for
input states taken from the $X$, $Y$, and $Z$ bases.
See the text for a more detailed explanation of the four
types of operations indicated above.
}
\end{table}

In order to obtain a classical non-local fidelity by local measurements,
it is necessary to verify the correlations of the output spin
components $X$, $Y$, and $Z$ in three separate measurements.
Each of these measurements corresponds to a classical local fidelity.
However, since only the correlation is verified, there are two
correct outcomes for each input, e.g. in the case given by equation
(\ref{eq:entangle}) above,
\begin{eqnarray}
F_{xz \to zz} &=& \frac{1}{4} \big(p_{xz|zz}(00|00) + p_{xz|zz}(00|11)
\nonumber \\
&& \hspace{0.1cm} + \; p_{xz|zz}(01|01) + p_{xz|zz}(01|10)
\nonumber \\
&& \hspace{0.1cm} + \; p_{xz|zz}(10|00) + p_{xz|zz}(10|11)
\nonumber \\
&& \hspace{0.1cm} + \; p_{xz|zz}(11|01) + p_{xz|zz}(10|10)
\big).
\nonumber \\
\end{eqnarray}
Using the classical local fidelities for $XZ$ to $ZZ$, $XZ$ to $XX$, and
$XZ$ to $YY$, the total non-local fidelity $F_{xz\to \mbox{\small ent.}}$ for
the characteristic observable operation given in equation (\ref{eq:entangle})
can be determined by
\begin{equation}
F_{xz\to \mbox{\small ent.}} = \frac{1}{2} \left(
F_{xz \to xx} + F_{xz \to yy} + F_{xz \to zz} -1 \right).
\end{equation}
Entanglement generation can thus be verified by a set of three
local fidelities.

\section{Complementary operations and process fidelity estimates}

As the discussion above illustrates, it is possible to characterize the
complete quantum coherent operation of a gate in terms of a
set of classical operations defined by different input state selections.
The effects of quantum coherence are then expressed entirely in terms of
directly observable classical input-output relations.
In fact, it is possible to uniquely identify a specific unitary
transformation using only two characteristic observable operations.
The condition for selecting these two operations is that
each input state of operation A must overlap with each
input state of operation B. The precise output states
of operation B then depend on the phase relations
between the output states of operation A, allowing
a complete test of the quantum coherence described by
$\hat{U}_0$ \cite{Hof04}.

Optimal sensitivity to quantum coherent
effects is obtained if the squared overlap of the input states
$\mid n \rangle$ and the input states $\mid k \rangle$
is equal to $1/N$ for any combination of $n$ and $k$,
e.g.
\begin{equation}
\mid k \rangle = \frac{1}{\sqrt{N}} \sum_{n=1}^N \exp[-i \frac{2 \pi}{N} k n]
\mid n \rangle.
\end{equation}
The output states of the characteristic operation defined by
the input basis $\{\mid k \rangle \}$ are then given by
\begin{equation}
\mid g_k \rangle = \hat{U}_0 \mid k \rangle =
\frac{1}{\sqrt{N}} \sum_{n=1}^N \exp[-i \frac{2 \pi}{N} k n]
\mid f_n \rangle.
\end{equation}
Any phase error in the output state components $\mid f_n \rangle$
will reduce the fidelity of this output. With respect to the
output components $\mid f_n \rangle$, the observable operation
$n \to f_n$ thus tests the output amplitudes, while the observable
operation $k \to g_k$ tests the phases. Since the unitary operation
$\hat{U}_0$ is completely defined by equation (\ref{eq:basic}),
a verification of both the amplitudes and the phases of the
output components $\mid f_n \rangle$ constitutes unambiguous
proof that the operation performed is actually $\hat{U}_0$.

As has been discussed in more detail elsewhere \cite{Hof04},
it is possible to obtain a reliable estimate of the overall
process fidelity $F_{\mbox{\small process}}$, defined as the overlap
of the ideal process matrix with the process matrix of the
experimental realization \cite{Bri04,Gil04}, using only the two
complementary classical fidelities $F_n$ and $F_k$,
\begin{eqnarray}
\label{eq:estimate}
F_{\mbox{\small process}} &\leq& \mbox{Min}\{F_n,F_k\}
\nonumber \\
F_{\mbox{\small process}} &\geq& F_n + F_k -1.
\end{eqnarray}
This result can be explained quite intuitively if one assumes
that the process fidelity corresponds to the probability of
performing the correct quantum operation. Since performance of
the correct quantum operation automatically implies that
every characteristic operation is correctly performed, all
classical fidelities must be equal to or larger than the
process fidelity. However, the classical fidelities also
include the probabilities of error processes. Such processes
can reliably perform one of the two complementary processes,
thus contributing to either $F_n$ or $F_k$. However, they
cannot perform both operations at once, so the maximal average
fidelity for error processes is $(F_n+F_k)/2=1/2$. The minimal
process fidelity is thus obtained by assuming that all error
processes have this maximal average fidelity of $1/2$.

It is now possible to apply this evaluation to the characteristic
operations of the quantum controlled-NOT gate shown in table
\ref{qcnot}. Complementary pairs are obtained by choosing any
pair of operations that has a different input input basis for
both the control and the target (that is, any pair that is
neither in the same line nor in the same column of the table).
Experimentally, it may be most convenient to choose a pair
of complementary operations that can be verified by local
measurements. This condition is fulfilled by pairs
consisting of a controlled-NOT and a reverse controlled-NOT operation,
e.g. the operation in the computational basis $ZZ$ and the operation
in the $XX$ basis. For this example, the estimate of the
process fidelity is
\begin{eqnarray}
F_{\mbox{\small process}} &\leq& \mbox{Min}\{F_{zz \to zz},F_{xx \to xx}\}
\nonumber \\
F_{\mbox{\small process}} &\geq& F_{zz \to zz} + F_{xx \to xx} - 1.
\end{eqnarray}
It is thus possible to evaluate the performance of a quantum
controlled-NOT by obtaining the fidelities of two classical
controlled-NOT operations performed by the gate.

Since it is usually not too difficult to change the local settings
for input and output states, the fidelity estimate can be optimized
by measuring the classical fidelities of all four controlled-NOT and
reverse controlled-NOT operations. The best estimate is then
obtained by using the lowest fidelity for the lower bound and
the highest pair of fidelities for the upper bound,
\begin{eqnarray}
\label{eq:4f}
F_{\mbox{\small process}} \hspace{-0.2cm} & \leq & \hspace{-0.2cm}
\mbox{Min}\{F_{zz \to zz},F_{xx \to xx},F_{zy \to zy},F_{yx \to yx}\}
\nonumber \\[0.2cm]
F_{\mbox{\small process}} \hspace{-0.2cm} & \geq & \hspace{-0.2cm}
\mbox{Max}\{F_{zz \to zz}, F_{zy \to zy}\}
\nonumber \\ && \hspace{-0.2cm}
+ \;
\mbox{Max}\{F_{xx \to xx}, F_{yx \to yx}\} -1.
\end{eqnarray}
It is thus possible to obtain a fairly reliable evaluation of the
quantum controlled-NOT from local measurements of its four
classical controlled-NOT operations.

\section{Classical fidelities and entanglement capabilities}

One of the applications of the process fidelity is to provide
a lower limit for any classical fidelity of the quantum device.
A lower bound of the process fidelity is therefore also
a lower bound for the fidelities of all characteristic operation
of the device. The bound obtained for a pair of complementary
characteristic operations $n$ and $k$ can thus be
applied directly to predict the minimal fidelity of any
other characteristic operation $l$. According to equation
(\ref{eq:estimate}), this lower bound for all classical
fidelities $F_l$ then reads
\begin{equation}
F_l \geq F_n + F_k -1.
\end{equation}
The measurement of only two complementary fidelities thus
provides an estimate for the fidelities of all other characteristic
operations that the device could perform.

In the case of a quantum controlled-NOT, this means that an
estimate for the fidelities of entanglement generation can be
obtained by measuring only the local controlled-NOT and
reversed controlled-NOT operations. The entanglement capability of
the gate can thus be verified without actually generating
entanglement.
For example, measurements of the controlled-NOT operation
in the computational basis $ZZ$ and of the reverse controlled-NOT
operation in the $XX$ basis can be used to estimate the
potential for generating entanglement from inputs in the $YY$ basis,
since
\begin{equation}
F_{yy\to \mbox{\small ent.}} \geq F_{zz \to zz} + F_{xx \to xx} - 1.
\end{equation}
Of course the same estimate applies to the fidelities of entanglement
generation from $XZ$, $YZ$, and $XY$ inputs.
A more precise estimate can be obtained from the four classical
local fidelities of the quantum controlled-NOT according to
equation (\ref{eq:4f}). This estimate reads
\begin{eqnarray}
\label{eq:fent}
F_{ij\to \mbox{\small ent.}} & \geq &
\mbox{Max}\{F_{zz \to zz}, F_{zy \to zy}\}
\nonumber \\ && + \;
\mbox{Max}\{F_{xx \to xx}, F_{yx \to yx}\} -1,
\nonumber \\
\end{eqnarray}
where the indices $ij$ indicate the four entanglement generating
input state selections $XZ$, $YZ$, $XY$, and $YY$.

Since the fidelities $F_{ij\to \mbox{\small ent.}}$ determine the
minimal probability of successfully generating a maximally
entangled state, it is possible to estimate the minimal
amount of entanglement that this fidelity can generate.
This estimate is based on the observation that the
concurrence $C$ measuring the entanglement of a mixed state
$\hat{\rho}$ can be estimated from the fidelity of a maximally
entangled state $\mid E_{\mbox{\small max}} \rangle$ using
\begin{equation}
C \geq 2 \langle E_{\mbox{\small max}} \mid \hat{\rho}
\mid E_{\mbox{\small max}} \rangle - 1.
\end{equation}
Since the minimal fidelity of obtaining a maximally entangled
state in the output of each entanglement generating operation
$ij$ is equal to the classical non-local fidelity
$F_{ij\to \mbox{\small ent.}}$ of the
operation, the corresponding estimate for the
concurrence $C_{\mbox{\small gate}}$ defining the
entanglement capability of the gate reads
\begin{equation}
C_{\mbox{\small gate}} \geq 2 F_{ij\to \mbox{\small ent.}} -1.
\end{equation}
With this estimate, the lower bound for the fidelity of entanglement
generation given by equation (\ref{eq:fent}) translates into
a lower bound for the entanglement capability given by
\begin{eqnarray}
\label{eq:cent}
C_{\mbox{\small gate}} & \geq &
2 \mbox{Max}\{F_{zz \to zz}, F_{zy \to zy}\}
\nonumber \\ && + \;
2 \mbox{Max}\{F_{xx \to xx}, F_{yx \to yx}\} - 3,
\nonumber \\
\end{eqnarray}
that is, the quantum gate can definitely generate entanglement if
the average of the maximal controlled-NOT fidelity and the maximal
reverse controlled-NOT fidelity is greater than $3/4$. In other
words, it is completely impossible that an experimental device
performs the classical local controlled-NOT and reverse controlled-NOT
operations at fidelities greater than $75 \%$ without also generating
corresponding am\-ounts of entanglement when operated in any of the four
entanglement generating operations.

\section{Conclusions}

In order to evaluate the performance of experimental quantum gate
operations, the essential properties of quatum gates have to be defined
in experimentally accessible terms. This goal can be achieved by
selecting representative sets of characteristic observable operations.
Each characteristic observable operation is given by
a set of distinguishable input states and their expected
distinguishable outputs. It is then possible to measure the
fidelity of each characteristic operation by comparing the actual
outputs with the expected ones, in close analogy to tests
conventionally performed on classical devices.

For the quantum controlled-NOT, a representative set of
characteristic operations is shown in table \ref{qcnot}.
These characteristic operations include the identity
operation, four classical controlled-NOT operations,
and four entanglement generating operations.
The latter operations cannot be verified in a single local
measurement, so it may be necessary to define their
fidelities in terms of three separate local fidelities
for the generation of correlations in $X$, $Y$, and $Z$.

The classical fidelities obtained for characteristic observable
operations allow estimates of the general quantum
properties of the gate. In particular, a lower bound for both
the process fidelity and the entanglement capability of an
experimental gate can already be obtained from only two
complementary classical fidelities. In the case of experimental
quantum controlled-NOT gates, it is therefore possible
to predict the amount of entanglement that a gate can generate
from the fidelities of two entirely local classical controlled-NOT
operations.

In conclusion, representative sets of characteristic observable
operations provide experimentally accessible evidence for the
successful implementation of quantum coherent gate operations.
Reliable estimates of the overall process fidelity and
the entanglement capability of the gate can be obtained efficiently
by combining the fidelities of two or more characteristic
operations. It is thus possible to perform efficient tests
of quantum gates using only a small number of well-defined
measurements.

\end{document}